\documentstyle[aps,prl,multicol,amsfonts,amssymb,epsfig]{revtex}
\renewcommand{\v}[1]{{\bbox #1}}

\newcommand{\sign}{{\rm sign}}

\newcommand{\rb}{{\bar{\rho}}}

\newcommand{\eq}{\begin{equation}}

\newcommand{\w}{{\omega}}
\newcommand{\zh}{{\hat{z}}}

\newcommand{\p}{{\partial}}
\newcommand{\gr}{{\nabla}}
\newcommand{\ra}{{\rightarrow}}
\def\eqa{\begin{eqnarray}}
\def\eea{\end{eqnarray}}

\newcommand{\be}{\begin{equation}}
\newcommand{\ee}{\end{equation}}
\newcommand{\ba}{\begin{eqnarray}}
\newcommand{\ea}{\end{eqnarray}}

\newcommand{\Eq}[1]{Eq.~(\ref{#1})}

\newcommand{\al}{\alpha}

\newcommand{\ga}{\gamma}
\newcommand{\Ga}{\Gamma}

\newcommand{\La}{\Lambda}
\newcommand{\om}{\omega}

\def\journal #1, #2, #3, 1#4#5#6{{\sl #1~}{\bf #2}, #3 (1#4#5#6) }
\def\pr{\journal Phys. Rev., }

\begin{document}
\draft
\widetext

\title{Edge tunneling in fractional quantum Hall regime}

\author{Dung-Hai Lee$^a$  and Xiao-Gang Wen$^b$ }
\address{
a) Department of Physics,
University of California at Berkeley, Berkeley, CA 94720, USA \\
b) Department of Physics,
Massachusetts Institute of Technology,
Cambridge, MA 02139, USA
}

\maketitle

\widetext
\begin{abstract}
\rightskip 54.8pt

We address the issue of an apparent disagreement between theory and experiment
on the I-V characteristic of electron tunneling from a metal to
the edge of a
two-dimensional electron gas in the fractional quantum Hall regime.
A part of our result is a theory for the edge of the half-filled Landau
level.
\end{abstract}


\begin{multicols}{2}

\narrowtext


Recently Grayson {\it et al}\cite{grayson} systematically studied the
current-voltage (I-V) characteristic for electron tunneling between a
metal and the edge of a two dimensional electron gas (2DEG) at fractional
filling factors ($\nu$).

They found that for $1/4\le\nu\le 1$,
$I\sim T^{\al-1}V$ for $k_BT>>eV$, and
$I\sim V^{\al}$ for $eV>>k_BT$, where
$\al\approx 1/\nu$. For primary filling factors such
as $\nu=1/3$, this result agrees
with the theoretical prediction based on the assumption of sharp edge
\cite{Wen}. However for other filling factors such as $\nu=\frac{n}{2n+1}$,
the experimental value ($\al\approx 1/\nu$) {\it disagrees} with the
sharp-edge prediction: $\al=3$. \cite{Wen} Perhaps what is even more
surprising is that scaling behavior in I-V was found even at filling
fractions between Hall plateaus, and at $\nu=1/2$.\cite{grayson,chang}
Several attempts has been made to explain this
result\cite{shaytov,conti,macdonald}.

The purpose of the present paper is to address the above disparity
between theory and experiment. However instead of dealing with an arbitrary
filling factor satisfying $\nu<1$, we shall limit ourselves to those correspond to
incompressible fractional quantum Hall states and $\nu=1/2$.
Our working assumption is that the edge modes can be described by a
quadratic action and for the most part we shall ignore dissipation.
\\

\noindent{\bf{The edge at incompressible filling factors:}}
\rm
Two facts at the edges of real samples may contribute to the
disagreement between experiment and theory.

1) The edge potentials in real samples are smooth. Moreover due to Coulomb
interaction the edge can ``reconstruct''. When that happens there are pairs
of left and right moving edge branches, and the exponent $\al$ becomes
non-universal because it depends on the interaction between these
counter-propagating
modes.\cite{CW} Although edge reconstruction can explain the departure of
$\al$ from the sharp-edge prediction, it cannot explain why $\al\approx
1/\nu$.

2) In reality charge and neutral modes have very different velocities. In
particular, for reconstructed edge the neutral-mode velocities $v_n$ are very
small due to the flatness of the effective potential. On the contrary the
charge velocity $v_c$ is of order $e^2/\hbar$ due the long range Coulomb
interaction.
In the
following we shall demonstrate that when temperature ($T$) and voltage ($V$)
fulfill the condition that $v_c\La_c>>k_BT,eV>>v_n\La_n$ (here $\La_{c,n}$
are the momentum cutoffs of the charge and neutral modes), the neutral modes
do not contribute to the I-V exponent $\al$. However, they are crucial in
restoring the proper short time behavior of the electron Greens function
($G_e(x,t)$) required by the electron Fermi statistics.
Due to the smallness of $v_n$, it is possible that Grayson {\it et al} are
probing the edge dynamics above the energy scale $v_n\La_n$. In that case
only charge mode contributes to the tunneling exponent, and hence
$\al=1/\nu$.

The momentum cutoffs $\La_{c,n}$ discussed above may have several
different origins. For example it can arise from 1) the fuzziness in the
position of tunneling events, and 2) the fact that above certain momentum the
edge modes become over-damped. (Toward the end of this paper we will
demonstrate that over-damping removes edge modes from  contributing to the
long-time decay of $G_e(x,t)$.) While
the first mechanism gives the same cutoff for charge and neutral modes, the
second can give $\La_n<<\La_c$. This is because for the charge mode there is an
extra conservation law - the charge conservation - that limits the number of
decaying channels. Indeed as was demonstrated experimentally, the charge mode
remains well defined at relatively high energies.

It has been pointed out in Ref.\cite{conti,macdonald} that
$\al=1/\nu$ can be obtained if one assumes that the tunneling charges are
added to the charge mode only. A difficulty with this explanation is that
for
$\nu^{-1}$ not equal to an odd integer this predicts an electron
Greens function that does not respect the electron Fermi statistics. The main purpose of the
the following
section is to demonstrate a mechanism by which only the charge mode contributes
to the long time decay of $G_e(x,t)$ (and hence $\al$), but both charge and
neutral modes contribute to ensure the correct short-time fermionic behavior of
$G_e(x,t)$.
\\

\noindent{\it{The two-mode model:}}

As far as electron tunneling is concerned it is sufficient to focus on only
two, a neutral and a charge, edge modes.
In the limit that the charge velocity ($v_c$) is much greater than all the
neutral ones ($v_n$) the action assumes a very simple form:
\ba
&&S=S_c+S_n\nonumber \\
&&S_c= \frac{1}{4\pi\nu}\int \frac{d\om dk}{(2\pi)^2}
k(-i\om + v_c k)|\phi_{k\om}|^2\nonumber \\
&&S_n=\frac{1}{4\pi\eta}\int \frac{d\om dk}{(2\pi)^2}
k(-i\om+v_n k)|\chi_{k\om}|^2.
\label{lkom}
\ea
Here $\phi$ and $\chi$ are the chiral fields associated with the charge and
neutral modes respectively.
(For purpose of regularization we shall keep $v_n$ and take the limit $v_n\ra
0$ at the end.)
The electron operator is given by
\be
\psi\sim e^{-\frac{i}{\nu}\phi}e^{-\frac{i}{\eta}\chi}.
\label{el}
\ee
The electron Fermi statistics requires
\be
e^{i\pi(\frac{1}{\nu}+\frac{1}{\eta})}=-1.
\label{elop}
\ee

The following is a sketch of how to obtain Eq.(\ref{lkom}) from the multi-mode
action for the edge of a fractional FQH state.\cite{Wen} First, we identify the
linear combination of the chiral fields corresponding to the total charge
displacement, and call it $\phi$.
Second, we identify a different linear combination $\chi$ so that the electron
annihilation operator can be written as Eq.(\ref{el}). Third, we integrate
out all the other linearly-independent chiral fields. In general the result
is rather complicated. In the limit that the energy associated with neutral
mode displacement is much smaller than the that of the charge mode we obtain
Eq.(\ref{lkom}).

Since the action in Eq.(\ref{lkom}) is quadratic, the  electron Greens function
is given by
\be
G_e(x,t)=e^{[G_{\phi}(x,t)-G_{\phi}(0,0)]/\nu^2}
e^{[G_{\chi}(x,t)-G_{\chi}(0,0)]/\eta^2},
\ee
where
\ba
G_\phi(x,t)&=&\int \frac{dkd\om}{(2\pi)^2}e^{i(kx-\om t)}
\frac{2\pi\nu}{k(-i\om + v_c k)}\nonumber \\
G_\chi(x,t)&=&\int \frac{dkd\om}{(2\pi)^2}e^{i(kx-\om t)}
\frac{2\pi\eta}{k(-i\om+v_n k)}.
\label{Gphi}
\ea
After performing the frequency integral and set $v_n=0$ we obtain
\ba
G_\phi(x,t)&=&2\pi\nu\int_0^{\infty}\frac{dk}{(2\pi)}\{
\frac{\cos{kz}}{k}+i\sign{(t)}\frac{\sin{kz}}{k}\}\nonumber \\
G_\chi(x,t)&=&2\pi\eta\int_0^{\infty}\frac{dk}{(2\pi)}\{
\frac{\cos{kx}}{k}+i\sign{(t)}\frac{\sin{kx}}{k}\}.
\label{ggphi}
\ea
In the above $z\equiv x+iv_ct$. First we note that
\ba
&&G_\phi(x,0^+)-G_\phi(x,0^-)=i\pi\nu\sign{(x)}\nonumber \\
&&G_\chi(x,0^+)-G_\chi(x,0^-)=i\pi\eta\sign{(x)}.
\label{sta}
\ea
It is Eq.(\ref{sta}) that ensures the correct short time
behavior of $G_e(x,t)$. By putting $x=0^+$ it is simple to show that
\ba
G_\phi(0^+,t)&\sim& e^{i\eta\frac{\pi}{2}\sign(t)}
e^{[G_{\phi}(x,t)-G_{\phi}(0,0)]/\nu^2}\nonumber \\
&\sim& e^{i\eta\frac{\pi}{2}\sign(t)}(\frac{1}{t})^{1/\nu}.
\label{gr}
\ea
Eq.(\ref{gr}) has the property that while its
short time behavior complies with the Fermi statistics, its
scaling exponent, $1/\nu$, is solely determined by the charge mode.

In reality with small but non-zero neutral mode velocities, we expect Eqs
(\ref{lkom}) and (\ref{gr}) to apply as long as
$(v_c\La_c)^{-1}<<|t|<<(v_n\La_n)^{-1}$, where
$\La_{c,n}$ stands for the momentum cutoffs of edge modes.
In terms of the experimental parameters it says that when
$v_c\La_c>>k_BT,eV>>v_n\La_n$ the I-V exponent $\al$ is $1/\nu$.
\\

\noindent{\bf{The edge at half-filling:}}
\rm
Now we turn to $\nu=1/2$ where gapless bulk excitations exist. In this
case even the meaning of edge mode is unclear.
In the following we spend some space to clarify this issue.

An electron has two basic attributes: its charge $e$ and its
Fermi statistics. In an
earlier work on $\nu=1/2$ one of us describes an electron as a boson carrying
charge $e$ and flux (statistical) $\phi_0$.\cite{lee} In this picture the
electron liquid is the superposition of a charge and a flux liquid. As an
electron tunnels in both liquids expand. The two edge
modes we shall encounter shortly are the edge deformation associated with these
two liquids. We note that in the bulk this description reduces to the recent
dipole theory of Refs.\cite{shankar,lee,haldane,read}.

In the limit of vanishing electron bare effective mass,
the low-energy action obtained by integrating out inter-Landau-level
excitations is given by\cite{lee}
\eqa
S&=&S_{int}[\frac{\gr\times\v a}{4\pi}]+
i\int d^2xdt\{\frac{1}{8\pi}\v a\times\dot{\v a}
+\v a\cdot\v j\nonumber \\
&&-\v b\cdot\v j-\frac{1}{4\pi}\v b\times\dot{\v b}\}.
\label{2}
\eea
In \Eq{2}: $\v j=\sum_i\dot{\v r}_i\delta(\v x-\v r_i)$ where
$\{\v r_j(t)\}$ are the {\it electron} coordinates;
$\zh\times\v a(\v x,t)$, $\zh\times\v b(\v x,t)$ are proportional to the
charge
and flux fluid displacements with $\frac{1}{4\pi}(\gr\times\v a,
\zh\times\dot{\v a})$ and $\frac{1}{2\pi}(\gr\times\v b, \zh\times\dot{\v b})$
being the induced
charge and flux 3-currents; the term $S_{int}$ is given by
\eqa
&&S_{int}[\delta\rho]=\int d^2xdt W(\v x)\delta\rho(\v x,t) \nonumber\\
&&\ \ +
\frac{1}{2}\int dtd^2xd^2x'V(\v x-\v x')\delta\rho(\v x,t)\delta\rho
(\v x',t),
\eea
where $W(\v x)$ is the confining potential and
$V(\v x-\v x')$ is the electron-electron interaction.
The partition function is given by
$Z=\int'D[\v r_j]D[\v a]D[\v b]exp\{-S\}$,
where $\int'$ denotes the Feynman path integral over $\{\v r_j\}$ and the
functional integral over $\v a$ and $\v b$ under the constraint
\eq
\frac{1}{4\pi}\gr\times\v a=j_0-\rb,\hspace{0.1in}
\frac{1}{2\pi}\gr\times\v b=j_0,
\label{22}
\ee
where $j_0=\sum_j\delta(\v x-\v r_j)$.

To separate the compressional and shear deformation of the charge and flux
liquids we write
\eqa
&&\v a=\v a_t+2\gr\phi\nonumber \\
&&\v b=\v b_t+\gr\chi,
\label{gauge}
\eea
where $\v a_t$ and $\v b_t$ are pure transverse vector fields. (
By construction
$\gr\phi$ and $\gr\chi$ do not affect the bulk charge and flux
densities.)
By substituting Eq.(\ref{gauge}) into Eq.(\ref{2}) we obtain
\ba
S&=&S_b+S_e+S_m+S_t\nonumber \\
S_b&=&S_{int}[\frac{\gr\times\v a_t}{4\pi}]+
i\int_{y>0} d^2xdt\{\frac{1}{8\pi}\v a_t\times\dot{\v a}_t
+\v a_t\cdot\v j \nonumber\\
&&-\v b_t\cdot\v j-\frac{1}{4\pi}\v b_t\times\dot{\v b}_t
\}\nonumber \\
S_e&=&\int dt
dx\{\frac{1}{2\pi}[i\p_t\phi\p_x\phi+v_c(\p_x\phi)^2]
-\frac{i}{4\pi}\p_t\chi\p_x\chi\} \nonumber \\
S_m&=& i\int dt dx (\hat{y}\cdot\v j_t)(2\phi-\chi)\nonumber \\
S_t&=&i\int d^2xdt (2\phi-\chi)(\p_{\mu}j_{\mu}),
\label{action}
\ea
where $\v j_t=\v j-\frac{1}{4\pi}\zh\times\dot{\v b}_t=\v
j-\frac{1}{2\pi}\zh\times\dot{\v a}_t$
is the transverse component of the bulk current $\v j$.
In obtaining \Eq{action} we have adopted the geometry that
the 2DEG occupies the space $y>0$. We note that since only
the charge displacement couples to the confining and Coulomb potentials, only
the charge mode acquires a non-zero edge velocity. Again for regularization
purposes we shall keep a neutral mode velocity $v_n$ and set it to zero at the
end.

In Eq.(\ref{action}) $S_m$ describes the mixing between the edge and bulk, and
$S_t$ describes the effect of tunneling on the edge displacements.
(In the following due to the consideration of
the electron Greens function we shall limit ourselves to a special tunneling
event where an electron is added (removed) at time $t=0$
($t=\tau$) at spatial position $\v x=0$. In that case
$\p_{\mu}j_{\mu}=\delta(\v x)[\delta(t)-\delta(t-\tau)]$.)
 By setting
$\delta S/\delta\phi=\delta
S/\delta\chi=0$ we obtain
\eqa
&&\frac{1}{2\pi}(i\p_t\p_x\phi+v_c\p_x\p_x\phi)=\delta(\v
x)[\delta(t)-\delta(t-\tau)]
+\hat{y}\cdot\v j_t\nonumber \\
&&\frac{1}{2\pi}(i\p_t\p_x\chi+v_c\p_x\p_x\chi)=\delta(\v
x)[\delta(t)-\delta(t-\tau)]
+\hat{y}\cdot\v j_t.\nonumber \\
&&
\label{anam}
\eea
Eq.(\ref{anam}) states that both tunneling and bulk transverse currents
contribute to expand
the charge and flux edge profiles. (The longitudinal bulk current causes
internal compression and hence does not contribute to expand the edge
profiles.)

By integrating out the bulk current fluctuations we obtain the following
effective theory for the edge
\eqa
S&=&S'+S_t\nonumber \\
S'&=&\int\frac{dkd\w}{(2\pi)^2}\{\frac{(-ik\w+v_ck^2)}{2\pi}
|\phi_{k\w}|^2+\frac{(ik\w+v_n k^2)}{4\pi} |\chi_{k\w}|^2\nonumber \\
\nonumber \\
&+&\frac{\ga(k,\w)}{4\pi}|\om||k||2\phi_{k\w}-
\chi_{k\w}|^2\}.
\label{damp}
\eea
In Eq.(\ref{damp})
\eq
\ga(k_x,\w)=\frac{1}{|k_x|}\int dk_y [\frac{k_x^2}{k_x^2+k_y^2}
\sigma_{tt}(\v k,\w)],
\label{kapa}
\ee
where $\sigma_{tt}(\v k,\w)$ is the bulk transverse conductivity.
The last term in Eq.(\ref{damp}) reflects the damping of the edge modes
by bulk excitations.
In the language of the two-mode model presented earlier $\phi$ and $\chi$
act as the charge and neutral mode respectively.
In the clean limit
$\sigma_{tt}(\v k,\w)\sim 1/|\v k|$ for $|\om|<<V_B|\v k|$. In that case
$\ga(k,\om)\sim 1/|k|$ and the last term in Eq.(\ref{damp}) changes the scaling
property of the electron Greens
function entirely. In the presence of disorder it is reasonable to assume that
$\sigma_{tt}(\v k,\w)$ saturates to a constant as $|\v k|\ra 0$. In that
case $\ga(k,\om)\rightarrow constant$ at long wavelength and the last term
of
Eq.(\ref{damp}) is a marginal perturbation. For simplicity in the following we shall assume that
$\ga(k,\om)=\ga_0\theta(v_B|k|-|\om|)$.

To compute the electron Greens function we calculate
\eq
G_e(\tau)=\frac{\int D[\phi]D[\chi] e^{-(S'+S_t)}}
{\int D[\phi]D[\chi] e^{-S'}}.
\label{gf}
\ee
In \Eq{gf}
\be
S_t=i\{[2\phi(0,0)-\chi(0,0)]-[2\phi(0,t)-\chi(0,t)]\}.
\label{st}
\ee
In the limit $\ga_0=0$ Eq.(\ref{gf}) gives
\eq
G_e(t)\sim e^{i\frac{\pi}{2}\sign(t)}\frac{1}{t^2}.
\label{free}
\ee
The above result has the long time behaviors implied by Grayson {\it
et al}'s experiment, and obeys $G_e(0^+)=-G_e(0^-)$ required by
the electron Fermi statistics.
For non-zero $\ga_0$ the calculation is a little more involved. To proceed we
define $\eta=2\phi-\chi$ and
$\zeta=2\phi+\chi$. After integrating out $\zeta$ we obtain
\eq
G_e(\tau)=\frac{\int D[\eta] e^{-(S_{\eta}+S_t)}}{
\int D[\eta] e^{-S_{\eta}}},
\ee
where
\ba
S_{\eta}&&=\frac{1}{4\pi}\int\frac{dkd\w}{(2\pi)^2}\{
A(k,\om)+\ga(k,\om)|\om||k|\}|\eta_{k\om}|^2\nonumber \\
S_t&&=i[\eta(0,0)-\eta(0,t)].
\label{efa}
\ea
In \Eq{efa} $A(k,\om)=[k(-i\om+v_ck)(i\om+
v_nk)]/(i\om+v_ck+2v_nk)$.
It is straightforward to show that
\ba
&&G_{\eta}(x,t)=G_{\eta}^R+iG_{\eta}^I\nonumber \\
&&G_{\eta}^R=-4\pi\int_{-\La}^{\La} \frac{dk}{2\pi}\int_0^{\infty}
\frac{d\om}{2\pi} S(k,\om) e^{-\om
|t|}\cos{kx}\nonumber \\
&&G_{\eta}^I=4\pi\sign{(t)}\int_{-\La}^{\La}\frac{dk}{2\pi}
\int_0^{\infty}\frac{d\om}{2\pi} S(k,\om) e^{-\om
|t|}\sin{kx},\nonumber \\
&&
\label{geta}
\ea
where
\eq
S(k,\om)=Im\{\frac{1}{A(k,i\om+0^+)+i\om |k|\ga(k,\om)}\}.
\ee
While $G_\eta^R$ is responsible for the long-time decay of
the electron Greens function, $G_\eta^I$ fixes up the short time statistics.
Unfortunately we will not be able to explicitly calculate the short time
behavior of $G_e$. This
is because such calculation requires knowledge of the bulk current-current
correlation function at both low and high frequencies. To compute the long time
behavior of $G_\eta^R$ it is save to set both $x$ and $v_n$ to zero.
We have evaluated $G_\eta^R$ in \Eq{geta} in the limit $\ga_0<<1$. The result
reduces to Eq.(\ref{free}) if
$v_B<v_c$, and it gives $G_\eta^R(t)=constant-(2+O(\ga_0^2))\ln|t|$ for
$v_B>v_c$.
Thus in the latter case the tunneling exponent receives an $O(\ga_0^2)$
correction. Nonetheless for
small $\ga_0$ (or when $\sigma_{tt}<<e^2/h$) such correction can be well within
the experimental resolution.

In conventional wisdom as the filling factor approaches
$\nu=1/2$ through the sequence $\nu=\frac{n}{2n+1}$ the number of edge mode
diverges. Therefore it is perhaps puzzling to see only two modes at $\nu=1/2$.
The two-mode model presented earlier precisely answers this question.
Indeed, \Eq{damp} can be viewed as the final product after infinite many
neutral modes are integrated out in the $n\ra\infty$ limit.
(In general we expect a similar two-mode effective theory to
be applicable to other compressible filling factors.)

An important omission in our discussion so far is the intrinsic and extrinsic
dissipation (such as those from mode-mode interactions and
phonons). Here we argue that dissipation can also prevent an edge mode from
contributing to the long time
decay of $G_e$. To make the point it is sufficient to consider the simplest
case where a single edge mode (described by chiral field $\phi$) is damped so
that its
chiral-boson action is given by
\eq
S_c=  \frac{1}{4\pi\nu}\int \frac{d\om dk}{(2\pi)^2}
k[-i\om + v k -i\sign(k)\Gamma_k]
|\phi_{k\om}|^2.
\ee
It is simple to show that the (real-time) Greens function is given by
\eqa
&&G_{\phi}^R(t)=2\pi\nu\int_0^{\La}\frac{dk}{2\pi k}
e^{-(ivk+\Ga_k)|t|}\cos{kx}\nonumber \\
&&G_{\phi}^I(t)=2\pi\nu\sign{(t)} \int_0^{\La}\frac{dk}{2\pi k}
e^{-(ivk+\Ga_k)|t|}\sin{kx}.
\label{lft}
\eea
\Eq{lft} suggests that while damping does not affect the short-time behavior
of $G_{\phi}$, it damps out the quantum fluctuation of $\phi$ at long
time \cite{CCN}
and hence
prevents it from contributing to the decay of $G_e$. Thus damping
can also remove the
contribution of neutral mode to $\al$.

To summarize, we find that there is an energy scale, $v_n\La_n$, above which
$I\sim V^{\al}$ with
$\al=1/\nu$. As the temperature/voltage is lowered below $v_n\La_n$ we expect
to see a change in $\al$ caused by
the neutral modes.

Finally we point out that under right condition the {\it
double-layer} $\nu=2$ state can exhibit $\al=1/2$.
However due to the spatial separation between the two edge modes and the
fact that only the outer edge couples sufficiently to the tunneling
electrode we do not expect the same for the single
layer $\nu=2$.

Acknowledgement: We thank C. Chamon, E. Fradkin and Y-C Kao for
useful discussions. DHL is supported in part by grands LDRD of the
Lawrence Berkeley National Laboratory and UCDRD of the
Los Alomos National Laboratory. XGW is supported
by NSF Grant No. DMR--97--14198 and by NSF-MRSEC Grant No.
DMR--94--00334. Both of us thank the Center for Theoretical
Sciences at National Tsinghua university, Taiwan for her
hospitality during our visits.


\end{multicols}

\end{document}